\documentclass[a4paper,9pt]{article}

\usepackage{INTERSPEECH2020}
\usepackage{multirow}
\usepackage{wrapfig}
\usepackage{CJKutf8}
\title{Cross-lingual Multispeaker Text-to-Speech \\ under Limited-Data Scenario}
\name{Zexin Cai$^{\star}$, Yaogen Yang$^{\dag}$, Ming Li$^{\star, \dag}$}

\address{
  $^{\star}$Electrial \& Computer Engineering, Duke University, Durham, NC, United States  \\
  $^{\dag}$ Data Science Research Center, Duke Kunshan University, Kunshan, China }
\email{ming.li369@duke.edu}

\begin{document}
\ninept 
\maketitle
\begin{abstract}
Modeling voices for multiple speakers and multiple languages in one text-to-speech system has been a challenge for a long time. This paper presents an extension on Tacotron2 to achieve bilingual multispeaker speech synthesis when there are limited data for each language. We achieve cross-lingual synthesis, including code-switching cases, between English and Mandarin for monolingual speakers. The two languages share the same phonemic representations for input, while the language attribute and the speaker identity are independently controlled by language tokens and speaker embeddings, respectively.  In addition, we investigate the model's performance on the cross-lingual synthesis, with and without a bilingual dataset during training. With the bilingual dataset, not only can the model generate high-fidelity speech for all speakers concerning the language they speak, but also can generate accented, yet fluent and intelligible speech for monolingual speakers regarding non-native language. For example, the Mandarin speaker can speak English fluently. Furthermore, the model trained with bilingual dataset is robust for code-switching text-to-speech, as shown in our results and provided samples.\footnote{https://caizexin.github.io/mlms-syn-samples/index.html}.  

\end{abstract}

\section{Introduction}
Recently, the combination of an encoder-decoder based text-to-spectrogram network and a neural vocoder has allowed machines to synthesize high-fidelity speech that is as natural as human. This technique can well equip text-to-speech (TTS) applications (e.g., audiobook reader, virtual assistants, navigation systems, etc.) in our daily life. However, these models, like Tacotron2 \cite{shen2018natural}, keep a certain level of limitations in controllability regarding latent speech attributes. Thus the models' robustness is limited and may be incapable of synthesizing speech with various speech characteristics. Then extensions on Tacotron2 have been proposed to address these problems: Yuxuan Wang et al. modeled the latent speech attributes by global style tokens (GSTs) while there are no explicit labels provided \cite{wang2018style}. Ye Jia et al. extend the Tacotron2 with conditioned features extracted from a speaker verification system to achieve speaker identity cloning and multispeaker TTS. \cite{jia2018transfer}.  

However, as bilinguists and multilinguists are commonly seen in today's world, the speech communication scenario becomes complicated. It is essential for speech analysis tools, including speech recognition and speech synthesis, to adapt this change for maintaining their current performance. The challenge is that languages, mostly, have different grapheme set and pronunciations between each other. This challenge motivates researchers to find and investigate shared representations between languages for speech analysis \cite{handbookguide, gales2015unicode, li2019bytes}. 

Even with appropriate representations for multiple languages, the model architecture needs to be upgraded in order to achieve multilingual processing for all speech analysis systems. For TTS, approaches are proposed for multilingual synthesis, even cross-lingual synthesis, based on classical statistical parametric speech synthesis (SPSS) \cite{li2016multi, ming2017light}. Since the end-to-end TTS models can generate speech with higher quality compared with classical methods, extensions on the end-to-end TTS frameworks also have been explored for multilingual modeling \cite{lee2018learning, zhang2019learning, zhou2020end, chen2019cross}. Normally, the voices of the multilingual TTS training datasets are different. Therefore, most TTS multilingual systems also support multispeaker synthesis. But the cross-lingual synthesis, where we can generate speech with foreign text for a monolingual speaker, is challenging. Yu Zhang et al. had achieved high-quality cross-lingual synthesis in a sufficient-data scenario \cite{zhang2019learning}. Zhaoyu Liu et al. investigated cross-lingual synthesis with limited data for each speaker, But the synthesized speech has moderate quality due to the data sparsity issue \cite{liu2019cross}. 

Motivated by the aforementioned works, in this paper, our focus is to achieve cross-lingual multispeaker TTS with limited data form two languages, English and Mandarin. We propose a model that incorporates speaker embedding and language embedding as the conditioned features for multilingual multispeaker TTS. The proposed model can generate high-quality speech for all speakers with respect to their own language. In addition, we investigate cross-lingual synthesis with the same model in a limited-data scenario by involving a bilingual TTS dataset. Results show that language-related knowledge can be transferred from the bilingual speaker to monolingual speakers, which enables us to generate fluent, high-fidelity, and intelligible speech in both Mandarin and English using monolingual speakers' voices.


\begin{figure*}[ht]
  \centering
  \includegraphics[scale=0.59]{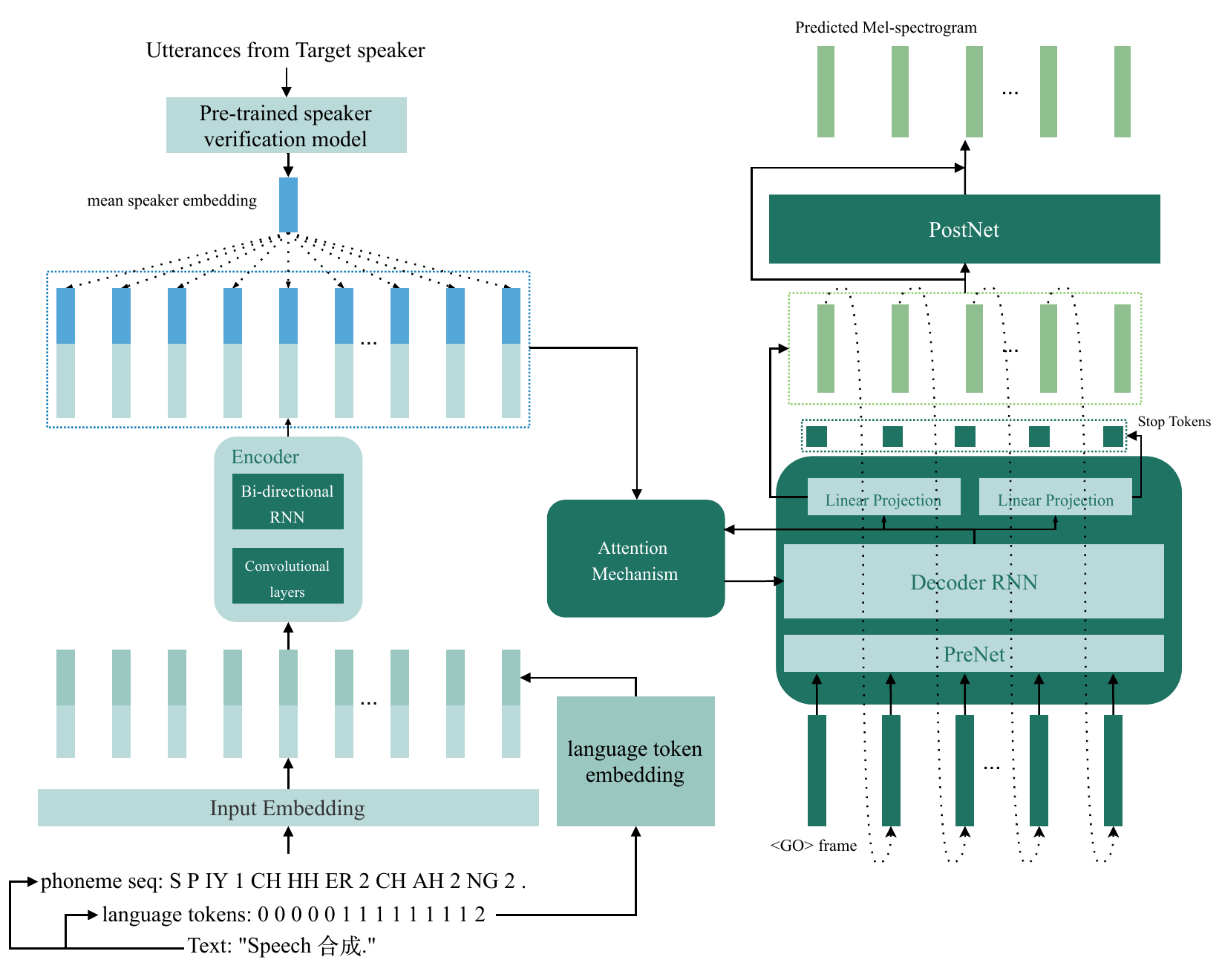}
  \caption{Proposed multilingual multispeaker TTS model}
  \label{fig:tts}
\end{figure*}

\section{Related works}
\label{sec:relatedwork}
Developing a multilingual multispeaker (MLMS) TTS model can relief the efforts of training multiple TTS models used for several voices with different languages. While the voice can be controlled by a text-independent speaker embedding in a multispeaker TTS system \cite{jia2018transfer, cooper2019zero},  TTS regarding multiple languages is more complicated due to different grapheme representations across languages. However, similar pronunciations between different languages can help reduce the gap of cross-lingual text-to-speech. Previously, Huaiping Ming et al. presents a light-weighted bilingual synthesis system that adopts concatenated vectors in the linguistic-feature level to manage two languages in one model. \cite{ming2017light} . Bo Li et al.  proposed an MLMS TTS approach based on conventional statistical parametric speech synthesis (SPSS) \cite{li2016multi}. They used the international pronunciation Alphabet (IPA) as the input representation and applied cluster adaptive language networks for generating the language-dependent linguistic features, followed by speaker-dependent output layers for different voices. 

Then in 2018, Bo Li et al. proposed a novel representation for all languages \cite{li2019bytes}. The representation, called Bytes, allows speech recognition models and speech synthesis models to manage multilingual processing. The performance of using Bytes in TTS is conducted and evaluated by another group of researchers \cite{zhang2019learning}. Experimental results in \cite{zhang2019learning} showed that phoneme inputs can achieve better performance than Bytes when used as the input for the MLMS TTS model. With sufficient training data (more than 500 hours), their proposed model is able to achieve cross-lingual synthesis with a high naturalness rate. The shared phoneme input is one of the keys to the cross-lingual synthesis, which is also stated in \cite{lee2018learning}. Similar pronunciations across languages result in close linguistic embedding vectors.  

Zhaoyu Liu et al. also used shared phoneme representation and extended the Tacotron2 by incorporating conditional embeddings for MLMS TTS \cite{liu2019cross}, which has a similar structure as our proposed model. However, we have the language-dependent Tacotron encoder designed for allowing the TTS model to synthesized code-switching text. Furthermore, we investigate MLMS TTS with limited data for each language and the performance in cross-lingual synthesis, while \cite{liu2019cross} investigate multilingual synthesis with limited data concerning each speaker. Xuehao Zhou et al. present a novel method to merge context information between languages by adopting word embedding from a pre-trained language model. Nevertheless, The cross-lingual synthesized speech has moderate quality, as shown in the figures from \cite{zhou2020end}.

\section{Method}
\label{sec:model}

\subsection{Input representation}
\label{subsec:rep}
 Code-switching is defined as more than one language occurring in one sentence or between sentences, either orally or in written form.  With the world's globalization, code-switching patterns in speech become a common case in many countries and regions. \cite{bernardo2005bilingual}. The language environment in the globalization inspires more and more bilinguists and multilinguists,  which motivates researchers to develop speech processing systems that can handle multilingual challenges. Furthermore, code-switching corpora have been collected and released for research related to speech communication in the recent decade \cite{lyu2010seame, shen2011cecos}, followed with various approaches proposed to address complicated speech analysis, including multilingual automatic speech recognition (ASR), language identification and language diarization with respect to multilingual scenario \cite{ahmed2012automatic, vu2012first, lyu2008language, lyu2013language}.  Likewise, TTS systems need to be improved for synthesizing natural speech for code-switching sentences \cite{zhou2020end}. 
 
 \begin{table*}[h]
\centering

\caption{Phonemes (without tone and stress) and their corresponding frequencies in LJ-Speech, DB-1 and DB-4}
\begin{tabular}{|c|c|c|c|c|c|c|c|c|c|c|c|}
\hline
Phoneme & LJS & DB-1  & DB-4    & Phoneme & LJS & DB-1   & DB-4   & Phoneme & LJS & DB-1   & DB-4   \\ \hline
J &  & 10088 & 12499 & X &  & 8050 & 11895 & Q &  & 5435 & 7489 \\ \hline
IY & 28587 & 54859 & 85601 & EH & 26397 & 3598 & 11791 & AA & 16976 & 11173 & 23205 \\ \hline
L & 32893 & 9420 & 23510 & AY & 12079 & 7479 & 15619 & UW & 15345 & 30630 & 44593 \\ \hline
SH & 7957 & 11456 & 17804 & OW & 10201 & 6921 & 13698 & Y & 4426 & 16540 & 27793 \\ \hline
N & 68392 & 33006 & 56359 & T & 65657 & 8698 & 26504 & JH & 4824 & 8994 & 13821 \\ \hline
AE & 21502 & 27640 & 42203 & NG & 7229 & 25895 & 36286 & AH & 102042 & 12558 & 33953 \\ \hline
G & 5901 & 6960 & 12298 & AW & 4248 & 9654 & 15397 & Z & 27845 & 5749 & 14135 \\ \hline
M & 23778 & 5967 & 14833 & AO & 16035 & 6970 & 14496 & S & 43700 & 5485 & 17965 \\ \hline
UH & 2856 & 7576 & 11253 & W & 20352 & 7151 & 15411 & CH & 4751 & 5118 & 7940 \\ \hline
D & 43601 & 14192 & 30390 & ER & 23525 & 15131 & 30264 & B & 15608 & 7577 & 15252 \\ \hline
F & 17018 & 4111 & 8890 & R & 40428 & 5025 & 16386 & K & 27866 & 3325 & 12650 \\ \hline
HH & 13785 & 7915 & 14745 & EY & 14695 & 4891 & 10838 & P & 20212 & 2496 & 8607 \\ \hline
V & 19628 &  & 4089 & DH & 29311 &  & 4716 & IH & 53904 &  & 11368 \\ \hline
TH & 3604 &  & 1250 & OY & 831 &  & 595 & ZH & 607 &  & 237 \\ \hline
   AX   &  156   &     & 418 &   &    &  &  &  &   &  &  \\ \hline
\end{tabular}
\label{tbl:phonemes}
\end{table*}
 
 However, one of the main challenges of code-switching TTS is that the grapheme set or the phoneme set between languages are different. Regarding that some phonetic pronunciations between different languages are close. Thus exploring a multilingual TTS model with minimum data requirement, including textual and vocal data, is possible and essential. Previous approaches, which are proposed for addressing multilingual issues in TTS, indicate that shared input representation across languages is one of the keys to realize cross-lingual synthesis \cite{li2019bytes, li2016multi, lee2018learning}. The shared representations include shared phoneme set, international pronunciation alphabet (IPA), and the Bytes coding \cite{li2019bytes}, where the phoneme representation can obtain better performance \cite{zhang2019learning}.  
 
 \begin{CJK*}{UTF8}{gbsn}
 In our work, we choose to use a shared phoneme set from CMU dictionary \cite{cmudict} to investigate bilingual multispeaker TTS and cross-lingual synthesis between Mandarin and English. As for Mandarin, the pronunciation representation called pinyin can be converted to CMU phoneme by the pinyin-to-cmu mapping table \cite{pinyin2cmu}. Since Mandarin is a tone-language, digits 1 to 6 are used to denote different tones, while `0', `1', `2' are used to mark the lexical stress for English. Although the tone and stress share the same annotations in our input, which may cause ambiguity, we have language identification tokens as another input stream. Moreover, language identification tokens are used to generate language-dependent encoding features while preserving the shared information between languages, like close pronunciations. Similarly, `0', `1', `2' are used for language identification in our input representations, where `0' represents the corresponding phoneme or stress annotation is from English, `1' is for Mandarin and `2' for language-unrelated symbols like punctuation marks. Take the phrase `speech 合成.' (speech synthesis.) as an example, two input sequences are obtained after the front-end text processing. One is the phoneme sequence `S P IY 1 CH HH ER 2 CH AH 2 NG 2 .', and the other is the corresponding language identification tokens `0 0 0 0 0 1 1 1 1 1 1 1 1 2' which has the same length as the phoneme sequence. We break up phonemes with its corresponding tones, e.g., `AH2' is converted to `AH 2', to allow our proposed model to share close pronunciations between Mandarin and English.  

\end{CJK*}

\subsection{Proposed model}
Our proposed bilingual multispeaker TTS model is illustrated in figure \ref{fig:tts}. The input text is converted into phoneme sequence and language token sequence, as introduced in section \ref{subsec:rep}. The phoneme sequence is converted to phoneme embedding sequence by a learnable lookup table. Correspondingly, the language tokens are converted to a 64-dimensional language embedding sequence through another learnable embedding table. Two embedding sequences are concatenated together as the input of the Tacotron encoder, which accumulates the linguistic and context characteristics of the input vector sequence with layers of convolutional layers and a bi-directional long short-term memory (BLSTM) layer. 

256-dimensional speaker embedding is concatenated with the encoder outputs for conditioning the network to synthesize expected voices. For the speaker embedding, we use the mean embedding derived from all embeddings extracted with a pre-trained speaker verification model \cite{cai2020fly} by feeding all training utterances of each speaker. We believe that it can induce the same performance as using a trainable lookup table yet costs less training time. Mel-spectrogram is used as the predicted acoustic feature in our bilingual multispeaker TTS model. Accordingly, we trained a neural vocoder, WaveRNN \cite{kalchbrenner2018efficient}, for converting the Mel-spectrogram back to audio signals. 

\section{Experiments}
\label{sec:exp}

\subsection{Datasets}
Our experiments are conducted with three TTS datasets, including the publicly available LJ Speech (LJS) dataset  \cite{ljspeech17} and two Chinese female voice datasets, DB-1 and DB-4, from Data Baker \footnote{https://www.data-baker.com/us.html} ( \textbf{LJS}, \textbf{DB-1} and \textbf{DB-4} are used as representations for both speaker identity and dataset in this section). DB-1 is publicly open, and DB-4 is a commercial one. LJS contains approximately 24 hours of audio-transcript English pairs recorded by a female English native speaker. The DB-1 has approximately 12 hours of Mandarin speech synthesis data recorded by a female Mandarin native speaker. The DB-4 is a bilingual dataset, which contains 12 hours of Chinese audio-transcript pairs, 6 hours of English pairs and 6 hours of code-switching data with a female Mandarin speaker.  

Table \ref{tbl:phonemes} illustrates the frequencies of all phonemes in three datasets. LJS contains only English utterances, while  DB-1 only Chinese utterances. Three consonants, `J', `X', and `Q' do not exist in the English dataset when using shared phoneme representations. However, these three phonemes frequently exist in the Mandarin dataset.  On the other hand, 7 phonemes are not presented in the Mandarin dataset while frequently existed in the English dataset, as shown in the table. The bilingual dataset DB-4 contains all phonemes. Most phonemes between two languages share the same representation in our experiments. This indicates that the intersecting shared phonemes may be less challenging to learn by a cross-lingual TTS system compared to those phonemes that only exist in one language. Moreover, the cross-lingual synthesis can be achieved when the model catches the pronunciation similarity of these phonemes between English and Mandarin.  

\begin{CJK*}{UTF8}{gbsn}
\begin{figure*}[ht]
  \centering
  \includegraphics[width=0.85\textwidth]{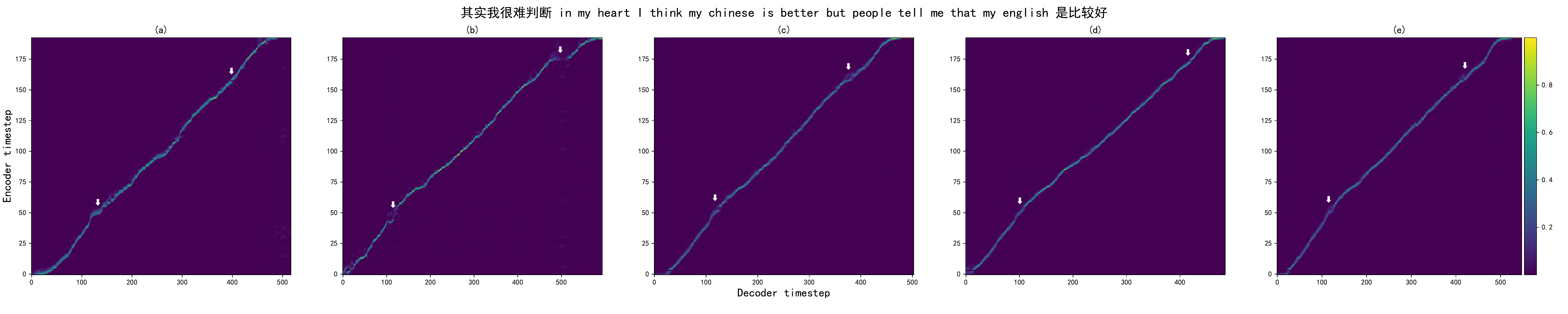}
  \caption{Attention alignments when synthesizing code-switching text `其实我很难判断\ in my heart I think my Chinese is better but people tell me that my English 是比较好' (Actually, it's hard for me to tell. In my heart, I think my Chinese is better, but people tell me that my English is better): (a) The alignment from BLLS with speaker DB-1; (b) The alignment from BLLS with speaker LJS; (c) The alignment from CLMS with speaker DB-1; (d) The alignment obtained from CLMS with speaker LJS; (e) The alignment obtained from CLMS with speaker DB-4;}
  \label{fig:ali}
\end{figure*}
\end{CJK*}

\subsection{Training setup}
We trained two bilingual multispeaker TTS systems with different datasets. The first system, notated by \textbf{BLMS}, is the bilingual multispeaker TTS model trained with DB-1 and LJS. The other system, notated by \textbf{CLMS}, is the system trained with all datasets, including the bi-lingual dataset DB-4. Although the latter system also can be used for bilingual multispeaker synthesis, we focus on its capability of cross-lingual synthesis here. All training audios are downsampled to 16 kHz. The vocoder WaveRNN is first pre-trained with the ground truth spectrogram-audio pairs from all three datasets. Then we finetune the pre-trained vocoder model with their ground truth alignment spectrograms after TTS training for each system.

\subsection{Objective evaluations}
The objective evaluation is done by speech synthesis MOS-scale rating, a categorical score from 1 to 5, with 0.5 increments. We ask 16 native Mandarin speakers (all speakers are familiar with English) to rate the synthesized speech concerning naturalness, similarity, and intelligibility. The naturalness is related to the quality of synthesized audios regardless of the content. The speaker similarity score is to measure how close is the synthesized voice to the expected speaker, while the intelligibility evaluates the clarity level of the speech content. We have three types of synthesized text for evaluating the performance, which are Mandarin sentences, English sentences, and code-switching sentences that contain both Mandarin and English content in each sentence. Each type of text has 15 sentences.

\begin{table}[]
\centering
\caption{The naturalness mean opinion scores (MOS)}
\begin{tabular}{|c|c|c|c|c|c|}
\hline
               & \multicolumn{2}{c|}{BLMS} & \multicolumn{3}{c|}{CLMS} \\ \hline
Text type        & DB-1       & LJS      & DB-1   & LJS  & DB-4  \\ \hline
Mandarin        &  4.08  &  3.07  &  3.94 & 3.10 & 4.11     \\ \hline
English        &  3.48  &  3.87  &  3.74 & 4.08 & 3.97     \\ \hline
Code-switching &  3.53  &  3.18  &  3.83 & 3.27 & 3.99    \\ \hline
\end{tabular}
\label{tbl:ntl}
\end{table}

The naturalness mean opinion scores (MOS) are shown in table \ref{tbl:ntl}. As shown in the table, the quality of synthesized audios reaches around 4.0, While the performance degrades when generating cross-lingual speech for monolingual speakers. For example, DB-1 obtains MOS with 4.12 when synthesizing Mandarin sentences but degrades to 3.64 for English sentences. As shown in table \ref{tbl:sml}, the speech synthesized by our proposed model can well preserve the speaker identity according to the speaker embedding. Most speaker similarity MOS are above 4, while scores lower than 4 can be observed in cross-lingual cases.  

\begin{table}[]
\centering
\caption{The speaker similarity mean opinion scores (MOS)}
\begin{tabular}{|c|c|c|c|c|c|}
\hline
               & \multicolumn{2}{c|}{BLMS} & \multicolumn{3}{c|}{CLMS} \\ \hline
Text type        & DB-1       & LJS      & DB-1   & LJS  & DB-4  \\ \hline
Mandarin        & 4.43  &  3.53  &  4.46 & 3.31 & 4.37     \\ \hline
English        & 3.61  &  4.04  &  4.18 & 4.20 & 4.17     \\ \hline
Code-switching & 3.81  &  3.52  &  4.28 & 3.38 & 4.22       \\ \hline
\end{tabular}
\label{tbl:sml}
\end{table}

\begin{table}[]
\centering
\caption{The intelligibility mean opinion scores (MOS)}
\begin{tabular}{|c|c|c|c|c|c|}
\hline
               & \multicolumn{2}{c|}{BLMS} & \multicolumn{3}{c|}{CLMS} \\ \hline
Text type        & DB-1       & LJS      & DB-1   & LJS  & DB-4  \\ \hline
Mandarin        4.76  &  1.90  &  4.72 & \textbf{3.53} & 4.81     \\ \hline
English        & 1.25  &  4.01  &  \textbf{3.86} & 4.56 & 4.47     \\ \hline
Code-switching & 2.60  &  2.34  &  4.18 & 3.84 & 4.59      \\ \hline
\end{tabular}
\label{tbl:igl}
\end{table}
The code-switching performance can be clearly observed from table \ref{tbl:igl}. Although BLMS can achieve bilingual multispeaker synthesis, the cross-lingual synthesis performance is poor, which matches the result in \cite{zhang2019learning}. The cross-lingual synthesized speech is unintelligible as the intelligibility MOS are below 2. However, while involving a bilingual dataset, CLMS is able to generate cross-lingual speech, even in code-switching cases, with intelligible pronunciations. Raters said that the synthesized speech is exactly like a foreign speaker speak another language with the accent from their native language. This indicates that, with our proposed model, using a bilingual dataset can significantly improve cross-lingual speech synthesis, although we only have limited data for each language. 

\begin{CJK*}{UTF8}{gbsn}
\subsection{Alignments}
In addition, the cross-lingual synthesis performance also can be seen from the attention alignments in  Figure \ref{fig:ali}. The synthesized content is a code-switching sentence. For system BLMS, we can observe clear breaks when the language switches in the sentence for monolingual speakers DB-1 and DB-4 in figure \ref{fig:ali} (a) and (b). However, the attention alignments obtained from CLMS are smooth even for monolingual speakers. This also implies that language-related knowledge can be transferred from the bilingual speaker to monolingual speakers with our proposed model. 
\end{CJK*}

\section{Conclusion}
\label{sec:conclu}
We present a bilingual multispeaker TTS approach based on shared phonemic representations. Our proposed model is able to achieve high-fidelity bilingual multispeaker TTS. In addition, results show that, by involving a bilingual dataset, the model is capable of cross-lingual synthesis, even for code-switching synthesis, under the limited-data scenario. We are able to obtain fluent, accented, and intelligible cross-lingual speech as monolingual speakers speak a foreign language. 

\footnotesize{\noindent\textbf{Acknowledgments} This research is funded in part by the National Natural Science Foundation of China (61773413) and Duke Kunshan University.\par}

\vfill\pagebreak

\bibliographystyle{IEEEtran}

\bibliography{mybib}

\end{document}